# Review of Cetacean's click detection algorithms

Mak Gracic, Guy Gubnisky and Roee Diamant

## I. ABSTRACT

The detection of echolocation clicks is key in understanding the intricate behaviors of cetaceans and monitoring their populations. Cetacean species relying on clicks for navigation, foraging and even communications are sperm whales (*Physeter macrocephalus*) and a variety of dolphin groups. Echolocation clicks are wideband signals of short duration that are often emitted in sequences of varying inter-click-intervals. While datasets and models for clicks exist, the detection and classification of clicks present a significant challenge, mostly due to the diversity of clicks' structures, overlapping signals from simultaneously emitting animals, and the abundance of noise transients from, for example, snapping shrimps and shipping cavitation noise. This paper provides a survey of the many detection and classification methodologies of clicks, ranging from 2002 to 2023. We divide the surveyed techniques into categories by their methodology. Specifically, feature analysis (e.g., phase, ICI and duration), frequency content, energy based detection, supervised and unsupervised machine learning, template matching and adaptive detection approaches. Also surveyed are open access platforms for click detections, and databases openly available for testing. Details of the method applied for each paper are given along with advantages and limitations, and for each category we analyze the remaining challenges. The paper also includes a performance comparison for several schemes over a shared database. Finally, we provide tables summarizing the existing detection schemes in terms of challenges address, methods, detection and classification tools applied, features used and applications.

## II. INTRODUCTION

Echolocation clicks are emitted by cetaceans for self-navigation or to locate prey [1]. In view of the high occurrence of echolocation clicks, these signals serve as important bioindicators that can be used to draw conclusions about the abundance of cetacean species. The analysis of these signals for presence detection or to classify individuals includes the temporal and spectral processing and the characterization of signals to investigate animal behavior patterns [2]. Indirectly, the detection and classification of clicks can serve as key techniques to understand anthropogenic impacts on the marine environment and to develop data-driven strategies and regulations. Since monitoring the activities of marine animals by passive acoustic monitoring (PAM) requires the analysis of large data sets, there is a need for automatic detection. The development of such detectors for echolocation clicks results from the broadband structure of these signals. While previous surveys are offered for detection of bioacoustics vocalizations [[3], [4], [5]], ours complements these by focusing on detection of transients, focusing on methods that work for these specific signals. We also present the databases used in the reviewed papers as well as implement most significant detection algorithms and compare them to the most commonly used detection software.

Echolocation clicks are impulse-like signals that are generated in the animal's nasal passage as a dictionary signal. To produce these signals, marine mammals push air through a pair of specialized organs called monkey lips or phonic lips [6]. The result of the air pressure passing through these lips is a "clapping" sound, often referred to as a *click*. The click sound can also be modified by a special organ in the animal's forehead that focuses the shape of the click signal, similar to an acoustic lens [7]. This process generates short transients that travel through the water and return to the animal as reflections. The animal uses these echoes to create a sound-based image of its surroundings. This last process involves the lower jaw bone, which receives the vibrations and then transmits them to the inner ear. From the sound-based images, the animal is able to analyze its distance to objects, the shape and density of reflectors, and even the speed and trajectory of potential prey [8]. Since we know for the most part how marine animals produce clicks, methods for recognizing such signals are offered for each individual species. Nevertheless, some general characteristics of clicks can be derived.

### A. Characteristics of Clicks

The structure of an echolocation click is characterized by its duration, frequency band, emission rate and directionality [9]. Clicks are typically short, pulse-like signals with a frequency band ranging from a few kHz in baleen whales to 160 kHz in some toothed whale species such as the harbor porpoise, depending on the species [10]. The duration of a click can range from microseconds to milliseconds [11], [12], and clicks are often produced in sequences: from a few clicks per second to several hundred [13], [14]. The direction and shape of the sound beam vary from a narrow beam of $5°$ in narwhals (Monodon monoceros) to a wide beam that is almost omnidirectional in sperm whales (Physeter macrocephalus) [15], [16]. The distinguishing features of the click usually include the bandwidth, the center frequency and the inter-click

M. Gracic, Guy Gubnisky and R. Diamant are with the Department of Marine Technologies, University of Haifa, Israel. R. Diamant is also with the Faculty of Electrical Engineering and Computing, University of Zagreb, Croatia. Corresponding author: Roee Diamant, email: roee.d@univ.haifa.ac.il.

This work was sponsored in part by the European Union's Horizon Europe programme under UWIN-LABUST project (project number 101086340), and by Project CETI via grants from Dalio Philanthropies and Ocean X; Sea Grape Foundation; Rosamund Zander/Hansjorg Wyss, Chris Anderson/Jacqueline Novogratz through The Audacious Project: a collaborative funding initiative housed at TED.

interval (ICI) [17]–[20]. The latter can change depending on factors such as water depth [21]. Differences in duration and pattern can also vary considerably; not only between species, but also between different individuals of the same species or even for the same individual under different conditions [18], [22], [23]. For example, it is known that the change in male sperm whales ICI for slow clicks is between 4 and 10 seconds [24]. The detection of clicks from a particular animal must therefore take into account the specific characteristics of the target clicks and distinguish them from clicks from other sources. Furthermore, to be robust, the detection scheme must be able to deal with sounds recorded from the marine environment, all of which may have transient characteristics similar to clicks.

### B. Challenges for Click Detection

The main challenges in detecting echolocation clicks lie in avoiding false detections due to anthropogenic noise disturbances (e.g., cavitation noise from ships), biological sources (e.g., snapping shrimp noise (SSN)) and transients that follow the strong tail distribution of clicks at sea [25]. If the propeller turns fast enough, the low pressure areas of the propeller can fall below the vapor pressure and the seawater can boil at ambient temperatures. When the bubbles behind the propeller reach ambient pressure, they implode and large, transient sounds reminiscent of bubble cavitation are emitted [26]. These signals are generated with an intensity of up to 180 dB 1$\mu$Pa/Hz@1m [27], which can be heard from tens of kilometers away. The SSN signals, in turn, are generated when a snapping shrimp closes its claws quickly. This creates a jet of water that is forced out between the claws and cavitation bubbles are formed. The maximum measured signal strength of SSN was found to be 220 dB (1$\mu$Pa/Hz@1m) [28]. Both cavitation and SSN, as well as transients, e.g., caused by waves, can easily be confused with the clicking of a whale [29].

An example of this can be seen in Fig. 1, where the time domain of a sperm whale click measured in the Bahamas (AUTEC data) (upper panel) is shown together with SSN clicks (bottom panel). Another challenge is the growing need to detect clicks in real time to enable a real-time system of fixed ocean observatories [30]. Here, a detector with low complexity is needed. In addition, echolocation clicks from multiple emitting animals may overlap in time due to the fast emission rate of the animals, which requires the ability to separate the sources. Finally, measuring the ICI poses another challenge as the sequence of clicks may change over time or overlap with other sources emitting at the same time. Considering the above challenges, a variety of techniques have been proposed to find a robust trade-off between detection and false alarm rate.

### C. Metrics of Performance Evaluation

For performance evaluation, common metrics are the probability of detection, the F-score and the Receiver Operating Characteristic (ROC). The probability of detection or sensitivity measures the ability of the detection method to correctly identify echolocation clicks. This can be within a certain buffer or for individual clicks. When the detection of individual clicks is of interest, e.g., for classification, the F-score is a balanced measure that combines precision (the proportion of detected clicks that are clicks) with recall (equivalent to the probability of detection). The ROC curve offers a compromise between the probability of detection and the false alarm. The Area Under the ROC Curve (AUC) is a measure of this trade-off, where 1 is perfect detection and 0.5 is the chance level. In the following, we present the available methods for detecting echolocation clicks in detail and comment on their suitability for different scenarios and signals.

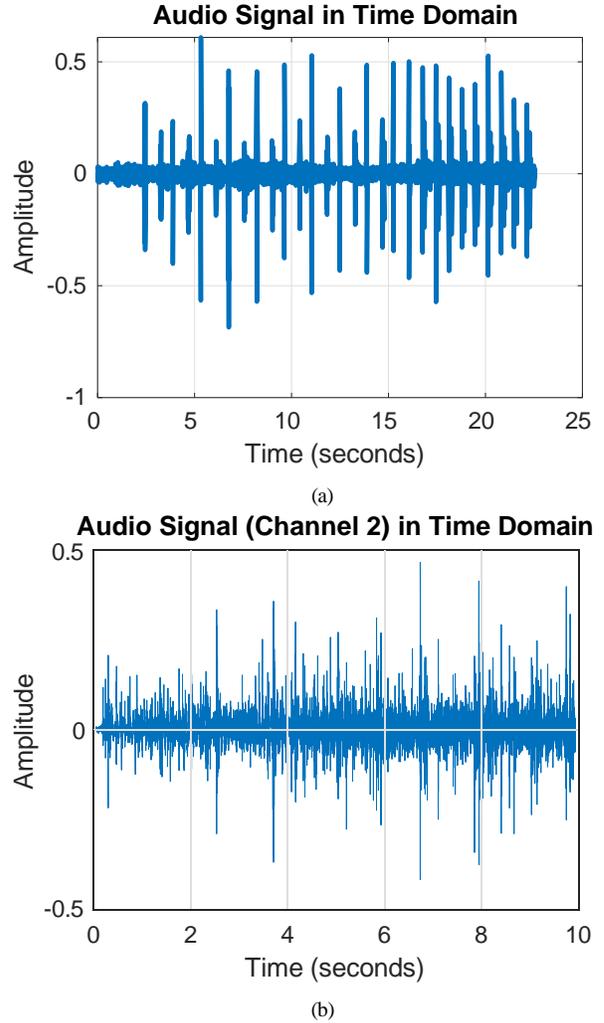

Fig. 1: Time domain of a sperm whale clicks (upper panel) and SSN transients (bottom panel).

### D. Structure of Survey

The Fig. 2 represents a structured breakdown of the click detection algorithms, divided into three primary methods. Three branches emanate from the root of the hierarchy. This subdivision represents different system models, ranging from knowing the actual signal structure of the click to no assumed information. The first branch, "feature analysis", uses the intrinsic properties of the signal, such as "phase", "frequency" and "energy", to distinguish echolocation clicks from signals originating from, for example, snapping shrimps. These techniques involve statistical analysis and thresholding, which makes them computationally efficient but prone to errors in detection due to their lack of adaptability to signal variation. Each of these attributes is analyzed by specific techniques such as wavelet transforms and spectral analysis for frequency or energy detection and the Teager–Kaiser (TK) energy operator for energy. The second branch, "machine learning based click detection", is proposed when no statistical information about the click is available. Based on a large dataset of labeled clicks, as well as noise samples, a model is trained to distinguish between clicks and non-click noises and to assign the detected click to its source. The focus is on "supervised convolutional neural networks (CNNs)", which are an important tool for pattern recognition in complex data sets. Other paradigms of neural networks and machine learning strategies, such as support vector machines, also fall into this category, indicating a variety of methods tailored to learn directly from data. The third and final branch, "data-dependent methods", uses a predefined knowledge of the expected analytical structure of the echolocation click to compare signals from the channel with a template. The aim is to recognize similarities between the signals and determine the detection based on predefined thresholds. Methods such as the "Tuned Filter", the "Side Test" and the "Adaptive Threshold" provide a means to improve the detection process. At the end of each category, we provide an overview of challenges that remain with each approach. Finally, a summary including useful information about the methods examined is presented. A detailed explanation of the implementation of the selected detector algorithms from each category, including the results obtained with real data, is presented. The algorithms were selected based on their relevance, ingenuity and impact on the field. A list of relevant available databases was then presented. Finally, all methods were grouped based on certain criteria to show some of the alternative criteria by which the methods could have been grouped and to highlight the similarity between the methods from the same groups. In addition, some common metrics are identified to evaluate the detection performance.

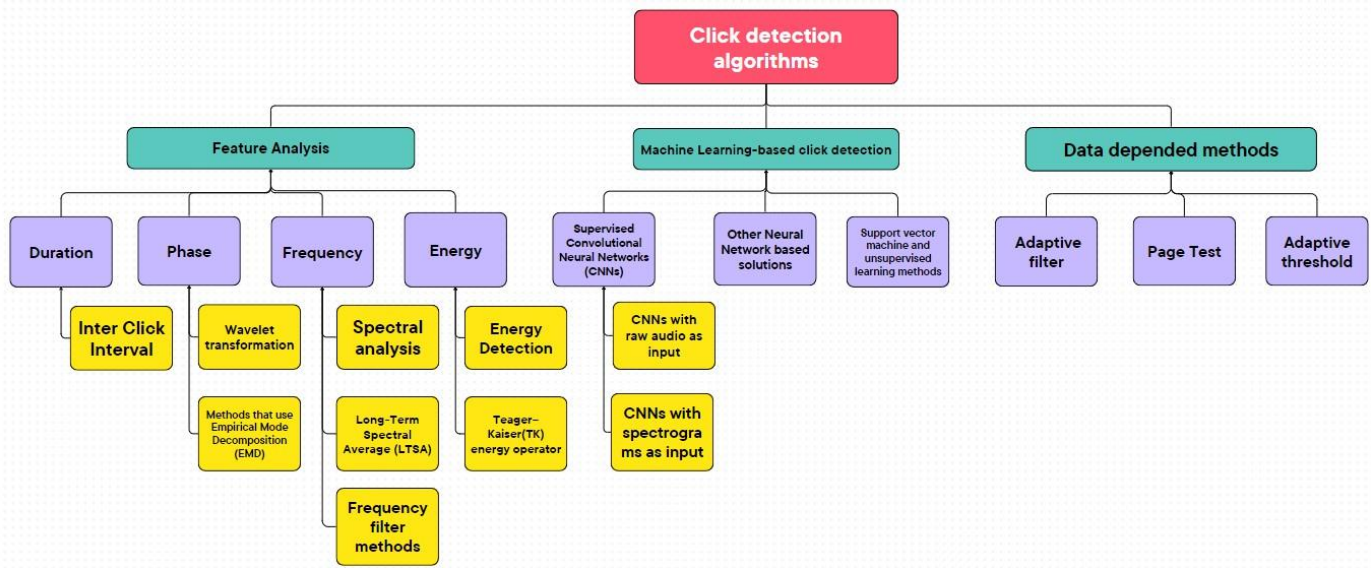

Fig. 2: The block diagram of algorithm categories

## III. FEATURE ANALYSIS

The term "features" refers to characteristics and properties of a signal that can be used to recognize or classify the signal. It is a process that involves the selection, extraction and evaluation of properties of the signal that are used to represent the structure of the click. The feature analysis approach to recognition focuses on isolating relevant attributes of the data through which key features are discovered, followed by recognition and classification. Below we provide an overview of the features used for click recognition.

### A. Duration

*1) Inter Pulse Interval (IPI):* The method described in [31] starts with an ultra-low power detector that performs an initial event analysis to significantly reduce false positives and thus increase reliability by ensuring that only probable clicks are forwarded for further analysis. A state machine analysis is then performed that integrates expert rules based on two important bioacoustic criteria: click duration and inter-click interval. The method uses the duration of the main click peak and the time between successive clicks to confirm the likelihood of a whale source. A click counter and validation process are an integral part of the system and provide further accuracy. This mechanism increments the click counter on pulses that match acoustic emissions from whales, comparing the duration between clicks with a reference interval to confirm click detection. The sensitivity of click detection is adjusted based on the observed rates of false positives and true positives. The strengths of this novel method lie in its energy efficiency and improved accuracy. The design also minimizes microcontroller activity, which significantly reduces power consumption. In [32], the authors present a novel method that focuses on the inter-pulse interval (IPI) of their clicks to improve the detection and classification of sperm whale vocalizations. In addition to amplifying the signals to improve SNR, the method also uses a phase-based IPI estimator to accurately recognize the inter-pulses. The method focuses on estimating the time between the major and minor pulses in a whale's click by using the phase-slope function (PSF) to accurately indicate the pulse positions and evaluating the IPI by the time difference between positive zero crossings. The method also includes feature extraction and segmentation to assess the consistency of the clicks and to separate valid IPIs from noise. The change from waveform-based detection to a phase characteristic-centred approach provides greater resilience to noise and signal distortion, although it relies on the assumption of consistent, multi-pulsed click patterns. An extension of this method can be found in [33], where the stability of the multi-pulse structure of identified transients is used to indicate the presence of sperm whales' clicks. The method starts with the transient detection phase using the TK- energy operator. For each detection, the multi-pulse structure (MPS) is calculated by taking the time interval between prominent pulses in the millisecond range. Assuming that the MPS representing the IPI of the whales or a multipath reflection is stable in time windows of a few seconds, a clustering solution is applied to find groups of clicks that fulfill the ICI (inter-click interval) conditions and whose variance of the MPS is below a certain threshold. This approach provides a robust solution for detecting sperm whale clicks in challenging environments, such as low SNR, a variety of noise transients, and simultaneously emitting whales. In addition, the method is computationally efficient and can be used in real-time applications. On the other hand, many valid clicks are overlooked to keep the false detection rate low, making the method unsuitable for individual click detection tasks.

*2) Inter Click Interval - (ICI):* Another method that uses the time difference as a recognition criterion is presented in [34] and focuses on the rhythmic characteristics of the click trains of beluga whales. It aims at detecting rhythmic pulse trains, separating click trains from multiple simultaneously clicking odontocetes and characterizing the ICI pattern. This approach handles subharmonics in the autocorrelation by rhythmic analysis. The multi-step algorithm starts by converting Time of Arrivals (TOAs) into a time-ICI map, then calculates a threshold to identify peaks corresponding to click trains, and then determines the threshold of the time-ICI map to create a binary map for analysis. This process leads to a detailed understanding of the rhythmic pattern over time. The authors also present the recognizability of a click sequence and the minimum ICI ratio required to separate two interleaved click sequences. The strength of this method lies in its robustness to the overlapping and mixing of click sequences from multiple sources. It efficiently distinguishes between individual click sequences embedded in a complex underwater acoustic environment. However, it assumes a rhythmic pattern of odontocetes clicks that may not cover all variations in acoustic emissions from cetaceans.

We see a similar approach in [35], where a method for recognizing burst pulses that resemble click-like events with a certain ICI is presented, which is the key to their identification. The method introduces the shift autocorrelation method (Shift-ACF), a novel approach that emphasizes repetitive events within an input signal to estimate the ICI, and is shown to be particularly effective in impulsive noise environments where conventional methods may struggle. The method is compared to the classic Cepstrum method, a frequency domain approach traditionally used for period estimation. While Cepstrum is effective in identifying temporal trajectories in a time-lag representation, Shift-ACF outperforms this method in impulsive noise environments and provides superior detection performance of burst pulses. Shift-ACF significantly improves detection performance in impulsive noise compared to the Cepstrum method, while Cepstrum performs better in Gaussian noise and low signal-to-noise ratio. However, the dependence of the Shift-ACF method on an accurate estimate of the ICI imposes limitations, particularly in the detection of burst pulses with highly variable ICIs. The method assumes that burst pulses consist of sequences of click-like events with a reasonably consistent repetition rate, which may not be universally applicable. Shift-ACF offers a more robust approach to background noise and reduces false alarms, increasing accuracy and reliability.

The study in [36] presents a method for offline detecting and classifying sperm whale echolocation signals using ICI characteristics. The method relies on an adaptive detection threshold adjusted to the ambient noise level. For detected regions of interests, the ICI and peak frequency are calculated and grouped into click sequences. Only click sequences with more than five signals of valid ICI pattern are considered for a second filtering that determines that the detected signals are valid clicks based on the peak frequency and duration. One of the main strengths of this approach is its adaptability to different acoustic environments due to the adaptive threshold, but the method relies much on thresholds for the ICI pattern and signal duration and spectra.

### B. Phase

The phase of the signal includes information about the temporal change of the signal. The phase is used in [37] to detect clicks by finding a zero crossing of the phase slope function of the signal. The phase slope function is a measure calculated by moving an analysis window over the signal and tracking the change in the slope of the phase spectrum at each shift. The derivative of the undistorted phase spectrum of the signal is calculated and indicates how the phase of the signal changes over time. By analyzing the slope of the phase spectrum, potential clicks are identified by finding the points where the function value changes from negative to positive. The authors also introduce the notion of centroid for clicks, i.e., the point at which the signal is "balanced" on the time axis, taking into account the phase or amplitude of the signal over time. This concept is valuable for tasks such as Time Difference of Arrival (TDOA) estimation, where the precise timing of these clicks is critical to determine the position of the source, and can be used as a reference point for multiple pulsed clicks, such as the regular clicks of sperm whales. Robustness to click source level and noise ratio is demonstrated using manually labelled data from regular beaked whale clicks and sperm whale clicks. The potential of phase jumps to represent a transient signal is also utilized by the wavelet transform.

*1) Wavelet transformations:* The wavelet transform involves the decomposition of a signal into its individual frequencies using small oscillatory functions that are localized in both time and frequency, the so-called *wavelets* - small waves that grow and decay in a limited period of time. The method in [38] combines the wavelet transform and a parameter called Short-Time Windowed Energy (STWE) to detect clicks. This parameter captures the unique shape of the click sounds that distinguishes them from other signals in the recordings and is calculated using

$$\text{STWE}_{WT}\,[s_k, lT_e] = \sum_{k-k_1} c_w\,[s_k, lT_c] \tag{1}$$

with $c_w$ the wavelet transform coefficient, $s_k$ the scale, $k_1$ and $k_2$ the scale range of the wavelet transform of the click, $T_e$ the sampling period and l, which defines the time resolution. First, the wavelet transform is performed over a specific buffer of potential clicks, followed by a calculation of the STWE parameter. The result is used to identify individual clicks by analyzing the peak of the STWE curves, which represents the exact time at which the sperm whale click was recorded, and the width of this peak, which correlates with the duration of the click. The ICI between identified clicks is used to verify detection and discard echoes. The results for both the simulation and the real collected data of sperm whale clicks show that the method is insensitive to noise transients. This method is then compared with a method that uses the Fourier transform instead of the wavelet transform. It is shown that the Fourier version of the method is less resistant to noise, particularly at low SNR. On the other hand, the properties of the STWE analysis should be adapted to the specific marine environment, and it is expected to be sensitive to changes in the structure of the clicks due to multipath effects. Such temporal changes in the structure of the click can be tracked by temporal modelling.

*2) Methods that use Empirical Mode Decomposition (EMD)*: Empirical mode decomposition (EMD) breaks down a signal into a series of eigenmode functions (IMFs) and is usually used to represent temporal variations in the signal. This empirical and adaptive process of decomposition takes the modes and frequencies present in the signal. Each IMF represents an oscillatory mode, and their accumulation encapsulates the information contained in the original signal. This temporal and spectral representation of the signal by its IMFs enables the isolation of broadband transient components, making EMD particularly effective for detecting non-stationary signals, such as clicks. This observation is utilized in [39], where the EMD is used for blind detection of clicks in a signal. An RMS ( Root-Mean-Square) window is then applied to each IMF to calculate an upper and lower envelope. The difference between these envelopes is then calculated and used to calculate the correlation coefficients between successive IMFs to assess similarity. A partial reconstruction of the signal influenced by the IMF with the highest correlation is then performed. Finally, a detection threshold is set based on a predetermined tolerance threshold and the partially reconstructed signal and any sample exceeding this threshold is identified, grouped and used for further analysis and classification. The classification algorithm calculates the strength of groups of samples that exceed a threshold and identifies the two groups with the highest strength as unique identifiers that are used to build an "EMD library" or IMF lookup table. These tables are then manually verified. A disadvantage of this method is that it works on the basis of the local characteristics of the signal rather than on a global basis that is uniform over time and frequency. Another method proposed in [40] additionally utilizes the estimation of the direction of arrival (DOA) of signal components for monitoring. The method is applied to a mixed model containing different signals that form the basis for DOA estimation. The individual signals are then isolated based on their unique characteristics. After extraction, the method performs endpoint detection on the signal components, using a "method of average energy". This process is crucial for identifying the exact start and end points of the signal components. SNR is also taken into account as it is critical to the clarity of the signal and the accuracy of analysis, such as DOA estimation, by measuring signal strength relative to background noise. The method uses EMD in combination with multi-layer adaptive decomposition, which increases computational complexity. The authors assume that the signals are oversampled or continuous, a condition that may not always be present in practical underwater environments. In [41] an upgrade is proposed, where a method combining the Complete Complementary Wavelet Ensemble Empirical Mode Decomposition with Adaptive Noise (CCWEEMDAN) and Power-Law Detector is presented. The method advanced beyond traditional EMD to handle modal aliasing and energy loss, which are particularly problematic for non-stationary, non-linear signals. The method includes iterative noise addition to improve scale continuity, wavelet decomposition to deal with noisy signals and EMD decomposition to extract residual components. The CCWEEMDAN method is combined with a power-law detector for transient signals, which analyzes the DFT sequence of the signal under two hypotheses - presence or absence of a signal in the midst of Gaussian noise. For this purpose, a non-parametric approach is used that analyzes the sum of squares of the power amplitudes of the DFT sequence. The method is shown to be effective in low signal-to-noise ratio scenarios, as demonstrated by simulated and real data. However, relying on iterative refinement and decomposition process, it leads to high computational complexity, which limits its application in practise. A time-frequency generalization of the EMD is the Hilbert-Huang transformation.

The method in [42] offers click analysis of rough-toothed dolphins. In this method, the raw acoustic data is first pre-processed to remove irrelevant low-frequency background noise. A Hilbert transform is then performed to create an energy envelope of the signal. An automatic click detector, focusing primarily on the ICI of echolocation clicks, incor-porates a strict SNR criterion and a careful peak detection algorithm, significantly reducing the number of false positives. The algorithm identifies potential click noise by looking for peaks in the energy envelope that meet certain criteria, including height and distance from other peaks, to distinguish them from random noise by checking the signal-to-noise ratio (SNR) to validate detection of echolocation clicks and not background noise.The strength of the approach lies in the rigorous assessment of the signal-to-noise ratio, which ensures the selection of potential echolocation signals. However, the method is sensitive to varying noise, since a uniformity of click characteristics is assumed. However, relying heavily on the SNR criteria can eliminate valid clicks, potentially underestimating the actual click rate. Furthermore, the assumption that clear peaks always represent single echolocation clicks may not hold true in scenarios with overlapping clicks or similarly loud sounds.For more dynamic environments, the Hilbert-Huang transformation (HHT) may be a solution.

The HHT process combines EMD and Hilbert spectral analysis (HSA). Specifically, the IMFs generated by EMD are used as input to HSA to obtain a time-frequency-energy representation of the signal, known as a Hilbert spectrum. Unlike the wavelet transform, the HHT does not require adjustment vectors for signal decomposition and is therefore considered more robust. By examining the Hilbert spectrum, transient echolocation clicks can be identified as components with concentrated, time-limited energy, characterized by their instantaneous frequency. In [43] the HHT is used to recognize sperm whale sounds. The clicks are identified by analyzing the first six modes of the Hilbert spectrum, arguably containing the key information of the click. A 'relevance/complexity' criterion is determined by calculating the ratio of the squared error between the original and the recovered signal (to the number of modes obtained) and used to evaluate the quality of the signal reconstruction. The paper discusses the advantages of using the HHT compared to the signal spectra. Next we discuss methods that focus on the latter analysis.

### C. Frequency

*1) Spectral analysis:* In spectral analysis, a signal is broken down into its fundamental frequency components in order to search for dominant features such as broadband transients. We distinguish between three feature types: spectral power, amplitude and phase spectrum. In [44], the text describes a three-tiered approach to classifying dolphin echolocation clicks: the supraspecies tier distinguishes based on the presence or absence of spectral peaks and notches; the second tier, the species tier, categorizes based on the frequency values of these peaks and notches; and the subspecies tier distinguishes two unique click types within Pacific white-sided dolphins. The first step of the click detection algorithm identifies potential clicks in the frequency domain using a fast Fourier transform (FFT) with spectral mean subtraction. The candidates were selected based on specific frequency and amplitude criteria. In the second step, the identified candidates were analyzed in more detail in the time domain. A high-pass filter and the Teager energy operator (explained in III-D1) were used to track energy peaks indicative of clicks. The strongest click noises within a given time frame were selected for further analysis. The spectral characteristics of the click sounds are then quantified with another FFT. The noise spectra are averaged and a subtraction of the spectral averages is applied to isolate the click spectrum, followed by statistical analysis to characterize the clicks of each species. To evaluate the utility of spectral features of clicks for classifying data, long-term spectral averages were examined for distinct patterns. The method was tested for recognizing and classifying the clicks of five dolphin species. However, recordings from the surveys were only included if they were single species schools and were excluded if other species were detected within 3 km or could not be identified due to low SNR. Handling multiple sources, in [30], spectral analysis is used to distinguish between the clicking sounds of sperm whales and the impulsive cavitation sounds of ships. After initial energy-based thresholding, spectral features are extracted from the potential click. Five statistical measures — mean, standard deviation, skewness, kurtosis and a normalized Shannon entropy — are used to analyze the features followed by a feed-forward neural network with a hidden layer of radial basis function units. And a logistic output function is used to classify the impulses into two categories: sperm whale clicks and ship sounds.

Processing gain is expected when combining spectral and temporal analysis. A joint spectral and temporal analysis is used to classify clicks in [45]. First, Fourier transforms of signal frames are observed to identify clicks with high SNR. Echolocation clicks are then identified based on their Teager energy, with noise level estimation and region magnification techniques to determine the start and end of the click. Clicks that were too close together are considered reflections. The cepstrum of each potential click is calculated to obtain a low-dimensional representation of the signal. Only the cepstral coefficients from 1 to 14 were used for classification, as higher order coefficients did not necessarily improve classification performance. Finally, the acoustic data of each species is modelled with a 16-fold mixed Gaussian Mixture Model (GMM) for classification. This approach allows the modelling of complex spectra with few data points. Spectral information can also be used through long-term analysis to detect periodicities in the signal.

*2) Long-Term Spectral Average (LTSA):* Long-Term Spectral Average (LTSA) is used to detect sporadic or rare biological sounds by identifying patterns, recurring events or anomalies in the frequency range of the signal. The LTSA visualisation calculates the spectral average of acoustic signals over longer periods of time, identifying patterns, trends and anomalies that differ from the surrounding sounds. In the context of click detection, LTSA can help to recognize recurring patterns, such as trains of clicks.

The method in [46], uses LTSAs from averaged sound pressure levels with specific frequency bins. In this semi-automated process, energy detection criteria are used to identify impulsive signals within a sampling window centred on the peak. The inter-click intervals (ICIs) between these detections are estimated, and signals with peak frequencies at bounded intervals are considered. These are then classified using an unsupervised learning method in which similar spectral shapes and ICIs are grouped within 5-minute bins and across time using clustering. For each group member, parameters such as click duration, ICI, spectrum, peak and center frequency and bandwidth are thresholded. Click duration was estimated by fitting an envelope to the absolute value of the waveform in the sample window. A combination of manual and automated analysis is also offered in [47]. The process involves the operation of the Triton software [48]. The signals were characterized by features such as long duration, stable interpulse intervals (IPI) and frequency modulation. The LTSAs were calculated for visual analysis. To facilitate manual analysis for the case of beaked whale type FM echolocation pulses, the echolocation pulses were sorted by peak frequency and peak-to-peak reception level to display high-quality signals.

## D. Energy

The energy of a signal can be used for detection based on power threshold or high order statistics.

*1) Energy detection:* In [49], an algorithm for eliminating multipath effects from sperm whale click sequences received from a single sensor is proposed. First, the clicks are detected using a moving average to find local maxima above a certain threshold. The study also included an analysis of the ICIs. The median ICI was calculated, with variations in ICI reflecting different behaviors or states of the whales. The consistency of ICIs over the entire click series was also analyzed. Next, a click separation algorithm is presented to identify and pair clicks. Potential click pairs are selected by time difference and SNR compatibility. Pairing is based on a similarity metric that uses statistical measures to determine whether or not two clicks are from the same click train. The algorithm uses Gaussian Mixture Models (GMM) for likelihood functions trained on validated click pairs for related clicks and random pairs for unrelated clicks. The similarity of the clicks is evaluated using features extracted from the clicks, including spectral and temporal information, which are categorized into three groups: spectral information, temporal information and inter-click interval (ICI) estimation. Feature selection aims to improve classification performance by adding the most informative features and reducing dimensionality. The method proceeds by finding the best subset of valid click pairs from all possible pairings. The clicks are then grouped into click trains, which are further categorized as direct path, surface path or reverberation. Gaussian Mixture Models (GMMs) were finally used to estimate the probability density function (PDF). Cross-correlation is performed to distinguish between direct and multipath click-trains at a sensor. Click trains that are assumed to originate from the same source but have different paths that show a high correlation are rejected as multipath. Click trains with a significant percentage of clicks identified as reverberation are also eliminated.

Energy is also for characterizing the click's structure. The method in [50] recognizes sperm whale clicks, where an adaptive threshold based on the median value of the total signal energy within a 5-minute recording is used to select potential clicks. The next phase involves cepstrum analysis, applied to both the amplitude and squared amplitude (energy) of the potential clicks to distinguish the stable interpulse interval (IPI) from the variable IPIs within the click structure. The average of the cepstral peaks identified within the delays is then calculated from at least 50 clicks within the same 5-minute recording. Similarly, in [51], the authors present a detection method that analyzes data across low, medium and high frequency bands using a short-time Fourier transform to reduce data size and align detection with expert analysis. The detection process calculates the spectral sum for each frequency band in each time window and identifies clicks as periodic peaks. By calculating the averages and standard deviations of these spectral sums over 10-minute intervals, the algorithm sets dynamic thresholds to distinguish potential sperm whale clicks from other sounds. The click detection criterion is considered to be met if the spectral sum exceeds a certain threshold in the low frequency band while remaining below the thresholds in the mid and high bands. The authors also focus on factors that influence the probability of detection, such as source level, directional loss, transmission loss and ambient noise level. An alternative way of calculating energy for transient detection is the Teager–Kaiser energy operator.

*2) Teager–Kaiser(TK) energy operator:* The Teager–Kaiser (TK) energy operator estimates the "mechanical" energy of the signal, which is a representation of the energy required to generate the signal. This estimate of the instantaneous energy of the signal is useful for detection because it provides insight into the dynamics and variability of the acoustic signal. The TK energy operator is particularly useful for detecting transient events such as clicks in recordings even in noisy environments. This is the case in [52], where detection of odontocete echolocation clicks of toothed whales is presented by developing an Energy Ratio Mapping Algorithm (ERMA). This scheme relies on species-specific features, such as increasing energy at certain frequencies. The ERMA scheme is used to create energy ratio maps for the target and non-target species. The study also describes the development of an energy ratio detector for suspected frequency bands identified by ERMA. A normalized Teager-Kaiser energy operator is then applied to the series of energy ratios to detect transients. Due to the high false positive rate, it is proposed to use ERMA as the first step in a two-step detection process, with a more sophisticated classifier as the second step to reduce the computational load. For the detection of clicks, a dynamically calculated threshold adapted to the noise is used. In [53] the TK energy operator is applied to analyze the given signal. The algorithm attempts to detect sperm whale clicks by identifying p0 and p1 pulses. To emphasise the click sounds, a matching filter is applied. This can prove challenging if the p0 pulse is much weaker than the p1 pulse, which can lead to detection errors as the algorithm is designed to recognize the highest peak within a click as the starting point. A skewness criterion is then applied to the output of the TK operator to help detect the presence of a click and avoid false positives. The length of the analysis window is one of the critical parts of this algorithm and a window size must be chosen that contains few click sounds, on the one hand, and is short enough to respond to rapid changes in click periodicity, on the other.

A forward-backward search is then performed over the peaks of the signal, separating them from all other signal values that have exceeded the threshold, with reference to the time of the highest peak, to locate the click. The forward and backward searches start at the highest peak and move forward and backward in time, respectively, until it reaches a point where the signal value falls below a certain threshold. It is assumed that the time interval between the two points contains the click sound. It has been shown that the same TC operator also works well under low SNR conditions (cf. [53]).

A similar pre-processing is performed with the acoustic analysis software developed by JASCO in [54], where three classification features are calculated: the number of zero crossings, the mean time between zero crossings and the slope of the time change between zero crossings. Since clicks of different species have different frequency components, the number of zero crossings can be a discriminating feature, while the mean time between zero crossings is related to the dominant frequency of the click sound. Since different species produce clicks at different frequencies, this measure helps to distinguish between these species-specific frequencies. The third feature represents the rate at which the time between zero crossings changes, which can be related to the frequency modulation of the click. The Mahalanobis distance metric is used to compare the features to a template created from manually labelled clicks. The choice of Mahalanobis distance is explained by its ability to account for the covariance between features.

The method in [55] detects echolocation pulses and buzz clicks by identifying peaks in the TKEO. The complete click sound, including the reverberation, is identified based on its energy profile. Accounting for the lower attenuation within the signal's lower frequencies, which can potentially distort the spectral characteristics of the signals, the median signal parameters are calculated using only the signals with the highest amplitude. The strength of this method lies in the combination of the broad spectrum of cross-correlation with the precision of TKEO. However, the reliance on manual scanning after initial detection could lead to human error or bias, and the efficiency of the method may be limited by the amount of data processed.

The combined works in [56] uses the TK operator as a preliminary step to enhance the signal and improve the SNR; The algorithm uses the phase slope function to detect the clicks and sets the length of the analysis window based on the average interval between clicks. The click sounds are detected by localizing the positive zero crossings of the phase slope function. Surprisingly, the structure of the clicks could also be detected when the phase slope function was applied directly to the non-optimised recording. Pre-detection based on the slope of the phase spectrum with respect to the center of the potential click. This center is calculated as the mean of the group delay function and a click is detected by searching for a positive zero crossing for the slope of the phase spectrum. The method requires statistics of at least one minute of recording. If more statistics are available, the high-potential machine learning can be adopted to recognize clicks.

## E. Remaining Challenges For Feature analysis

The variability of signal characteristics due to individual differences between marine mammals, such as different click patterns and vocalisation types, poses a major challenge for research. The algorithms must be adaptable enough to accurately analyze a wide range of signal types under different environmental conditions. Additionally, the presence of background noise, including natural and anthropogenic sources, complicates signal processing and feature extraction. The dynamic nature of underwater environments, characterized by changing temperature and salinity, affects the sound propagation. This variability requires algorithms that can adapt to such fluctuations. Another aspect is real-time processing, which is required for many applications, such as monitoring shipping traffic and identifying species for nature conservation. The development of algorithms that are accurate and efficient in real-time data processing remains a major challenge.

As for future research directions, the development of integrated solutions combining acoustic properties with oceanography information and acknowledge of the animal's activity (e.g., vocalizing only upon surfacing) could provide better detection results. Another potential research avenue could be the exploration of advanced signal processing techniques such as the Hilbert-Huang transform, which offers advantages in analyzing non-linear and non-stationary data prevalent in marine mammal acoustics.

## IV. MACHINE LEARNING-BASED CLICK DETECTION

Machine learning (ML) techniques have been proposed to capture the variability in the structure of the click by learning a model for a valid click from datasets containing such signals as well as noise and perturbation intensities. These techniques are known for their ability to process and analyze large amounts of data quickly and are useful for detecting patterns in the data. One ML technique that has proven useful for automatic click detection is the Multilayer Perceptron (MLP). The MLP approach is a type of feed forward artificial neural network (ANN) [57]. An MLP consists of at least three layers, including an input layer, at least one hidden layer and an output layer. Each of these layers is fully connected to the previous and subsequent layers. The weights of these connections are usually trained by backpropagation, an iterative supervised learning technique in which the differences between the given output and the desired output are calculated as an error and the calculated error is then used to determine the new weight [58]. In the context of click detection, MLP is useful due to its low computational cost, high performance and simple structure [59].

Convolutional neural networks (CNNs) are another type of ANN. In contrast to MLPs, the layers of the CNN are sparse. This benefits the generalization of the network, as overfitting is reduced. It also allows the network to focus on the important features of the input data while ignoring irrelevant or redundant information, which in turn leads to automatic feature learning from raw audio data without the need for manual feature extraction. A CNN is characterized by its convolutional layers and pooling layers. The former represents a set of kernels that learn and extract features from the input data and create the feature map that represents the presence or absence of a particular feature at each location in the input data. Pooling layers are often placed between the convolutional layers to reduce the spatial dimensions of the data. CNNs are considered parameter efficient and are better suited for recognizing spatial hierarchies than MLPs. This is achieved through a concept known as *local connectivity*, where each neuron is connected to its local region. This technique reduces the number of parameters by allowing different parts of the network to specialize in high-level features such as a texture or a repeating pattern [60]. For click detection, CNN offers the advantages associated with the small size of the network. While CNNs can handle spatial hierarchies in gridded data, the Recurrent Neural Network (RNN) is better suited to the task of analyzing sequential data sets such as time series with sampling dependencies. The reason for this is the ability of RNNs to recognize patterns in sequences and learn from them. RNNs maintain a hidden state from one step in the sequence to the next. In this way, they maintain a memory for previous inputs in their internal structure. This memory is used to recognize causality within the dataset and is therefore useful for applications such as speech recognition, natural language processing and video activity recognition. For click recognition, RNNs can use their memory to draw information from a series of clicks. One of the main problems in using RNNs is overcoming the vanishing gradient problem, a phenomenon that occurs during the training part of the network. In this case, the gradient approaches zero, which leads to a loss of information and makes it difficult for the network to learn and update its weights. A special type of RNNs that takes this problem into account are Long Short-Term Memory (LSTM) networks. In contrast to RNNs, LSTM networks are characterized by their gating mechanisms, namely input, forget and output gates. The use of these gates enables the network to remember or forget observation inputs, making it more resilient to the vanishing and exploding gradient problem. In the context of click detection, the LSTM can be useful by adaptively distinguishing between clicks and other linear impulse noises from spectrograms.

A simple but sometimes effective learning method is the Support Vector Machine (SVM). An SVM finds a hyperplane that best separates different classes of data with the maximum margin. The margin is defined by the data points (support vectors) that are closest to the decision boundary. The so-called kernel trick allows SVMs to support high-dimensional spaces, which is why they often use kernel functions to map input data into a higher-dimensional space. The main advantage of SVMs compared to other neural networks is the lower risk of overfitting, which is especially important when the training data is limited. This is particularly important when the training data is limited. For click detection, this is relevant when there are only a few acoustic recordings on which to develop a detector.

While SVMs focus on maximizing the marginal distribution, which is limited by their ability to set constraints for classification, the Gaussian Mixture Model (GMM) learning approach is an alternative for probabilistic modelling of data distributions.

Assuming that the data can be clustered into classes of Gaussian distributed samples, GMM aims to determine the distribution parameters of each class by likelihood maximization. The result can be applied to click detection by using GMMs to model the distribution of relevant extracted click features or to detect anomalies that differ from the "normal" distribution of clicks. The structure of GMMs offers a soft, probabilistic assignment of data points, allowing constraints to be set as part of the clustering process. This can be a restriction on the distribution parameters between classes, samples that must or must not be clustered together, and a minimum number of samples within the class. This proves useful for the detection of clicks by identifying and modelling background noise of the recording with GMMs, which increases the click-to-noise ratio. Another form of generalization model is the Generalized additive model (GAMs) , which develops a statistical model for the relationship between the input variables to represent the probability density function of the predictor's variables. This negates the need to create a single global model while handling non-linear relationships. For click detection, this is very handy as they can be used to find temporal patterns for the presence or absence of clicks.

In the following, we categorize the papers according to the classification into ML techniques, specifically *supervised convolutional neural networks*, *alternative supervised neural networks* and *unsupervised learning models*. The contributions are further categorized according to the type of input they are best suited for, the underlying architecture they use and their adaptability to click detection.

## A. Supervised Convolutional Neural Networks (CNNs)

*1) CNNs with raw audio as input:* Convolutional Neural Networks (CNNs) are deep learning models that recognize a non-linear hierarchical order in the features of a valid click. CNNs use layers of convolution to learn spatial hierarchies of features from input images. In the case of click detection, the inputs are usually raw temporal acoustic data or spectrograms. When CNNs are applied to spectrograms, they can detect patterns in both the time and frequency domains. Convolutional layers allow the network to focus on localized features, ensuring that slight variations or shifts in the position of the click in the spectrogram (relative to the start of the input) do not affect recognition accuracy. By progressively abstracting information through its layers, a CNN captures both the broader context of echolocation signals and the fine-grained details of a particular click. By applying these principles, CNNs have already been successfully used for click detection.

Since clicks are short signals, using one-dimensional audio signals as a base layer offers the CNN the opportunity to learn important features of the signals that distinguish them from noise, cavitation or SSN. The work in [61] uses CNNs to automatically detect echolocation clicks of odontocetes from acoustic data recordings. The proposed method involves two-step detection in which a deep CNN is trained on both synthetic and real data to discriminate between click and non-click clips at different SNR values. Subsequently, the trained CNN is converted into a full convolutional network to minimize computation time and overcome the restriction to fixed-size inputs. This approach enables fast data processing. An energy normalization procedure allows the management of variable input lengths. In post-processing, the authors use the Teager-Kaiser energy operator (TKEO) to search for a transient and then the Gabor curve fitting method to fit a discrete Gabor signal to the acoustic data of a click to obtain a more accurate time synchronisation of the start and end points of the click. The use of CNN has been further developed using the spectral representation of the signal.

The CNN architecture is well suited to recognizing patterns in grid-like topologies such as the spectrogram of audio signals. The work in [62] presents a comprehensive study on the use of Deep CNNs to recognize porpoise clicks from acoustic data. The authors investigate different CNN architectures and the performance of different CNN models on this task and compare the methods in terms of their accuracy. Six CNN architectures, including LeNet, LeNet variants and ResNet-18, are developed and tested on a dataset of bottlenose dolphin clicks. "Traditional" texture feature extraction classification is also explored. Both the spectrogram pixels and the extracted LBP features are used as input. The results show that CNN outperforms these methods for echolocation clicks belonging to one species. The article concludes that ResNet-18 performs the best of all the architectures tested. This can be explained by ResNet's ability to ignore layer connections that bypass one or more layers, ensuring low sensitivity to additional layers. Results of SVM and MLP classifiers are compared with raw pixel values of the spectrogram images to evaluate the effectiveness of CNNs. The success in recognizing the clicks is attributed to the distinct pattern that is evident in the time-frequency domain. For the manual analysis of sperm whale clicks in acoustic recordings, a customized annotation interface was used in [63] combining with a click detector and the calculation of arrival times, IPI and background noise levels. These metrics are used in [63] to analyze the behavior of sperm whales in the presence of anthropogenic noise. For the detection of clicks, spectrograms are used as input to the CNN to utilize the broadband characteristics of the signal as well as temporal features such as IPIs and ICIs.

*2) CNNs with spectrograms as input:* The spectrum of the signal enables the identification of stationary patterns in the signal. Using spectrogram images as the input to a CNN [64] explores these patterns to recognize cetacean vocalizations. These signals include clicks, whistle and whale song signals of different whale species. For performance evaluation, three types of measures are used: Original Test Data (OTD) that serves as a baseline to evaluate the effectiveness of the CNN under ideal conditions, Synthetic Test Data (STD), which tests the robustness and adaptability of the CNN model, and Practical Test Data (PTD), that evaluates the performance of the CNN in real-world conditions. The former is derived directly from the dataset; the STD is generated by artificially modifying OTD; and PTD is created to simulate real-world conditions by combining original whale sounds with oceanic ambient noise.

Together with the detection accuracy, precision, recall and F1-score, these metrics demonstrate the efficiency of the CNN model in detecting and classifying signals.

A combination of CNNs with other deep learning methods is demonstrated in [65]. The clicks of sperm whales are detected and classified using deep machine learning techniques. A CNN is used for click detection while recurrent LSTMs are used for classifying clicks into categorical types and to recognize dialects of vocal clans. In addition, the principal component analysis (PCA) and t-Distributed Stochastic Neighbor Embedding (t-SNE) algorithms are used to calibrate the models parameters. Transfer learning is used for training on codas from the Eastern Tropical Pacific (ETP) dataset.

### B. Other Neural Network based solutions

The detection of click sounds has been demonstrated using MLP, RNN and LSTM. The use of MLP is motivated by its success in speech recognition, environmental noise classification and seismic signal analysis. The method in [66] uses a combination of different MLPs and statistical analysis to distinguish between regular clicks, creaks or noises. The statistical features used as input are the standard deviation of the energy values within each time window and the dynamic range , which is defined by a ratio between the maximum level and the background noise level within the time window. The detection is performed over short time buffers of 2 seconds and can therefore be considered real-time, but the achievable misclassification rate is high. This could be due to the strong assumption that a large number of identical click structures exist in the time window analyzed for statistical accuracy. Another MLP-based approach can be found in [67]. The authors use the Chimp optimization Algorithm (ChOA) to train an artificial neural network and to set a fuzzy logic for parameter adjusting. The input is a pre-processed spectrogram from which the features are extracted by averaging the cepstral values and applying cepstral liftering. The control parameters of the ChOA algorithm are adjusted in three stages: Fuzzification, fuzzy inference and de-fuzzification. The method uses membership functions to convert the input into fuzzy sets. The results of the fuzzy inference are then converted into quantitative data using the defuzzification process using two membership functions. Comparison without Fuzzy logic as well as with the coronavirus optimization algorithm, Harris-Hawks optimization, the Black Widow optimization algorithm and a Kalman filter shows an advantage in both classification rate and convergence. However, the method performance depends on the quality of the input data. This can be avoided by utilizing the sequential properties of echolocation clicks to learn from high-dimensional data using the residual neural networks (ResNet).

The ResNet's ability to effectively learn hierarchical features makes it suitable for learning from image-like representations, such as spectrograms, so that it can exploit both temporal and spatial information. The ResNet model proposed in [68] is used for segmenting, recognizing and classifying audio segments as killer whale sounds or noise. The method used is a modification of the ResNet architecture. The data is divided by a sliding window into short segments that are used as input to the ResNet-based neural network. The network performs binary classification for presence detection to determine if the segment contains clicks. For evaluation metric, a measure for the time-based precision is offered to measure the accuracy of click detection over time. This is shown to be useful for generalization of time-dependent processes. Nevertheless, the performance is sensitive to the choice of detection threshold. To solve this robustness problem, data augmentation has been proposed.

Data augmentation is used to expand the database for training using simulated clicks. Data augmentation is used in [69], where the EfficientNet B0 model is used as the backbone of the CNN pipeline to distinguish between environmental noise, dolphin sounds, biological clicks and ship noise. This model scales the depth, width and resolution of the network for robust detection. The input for the CNN is a multi-channel spectrogram. Audio enhancement in the time domain includes time shifting, pitch shifting and changing the SNR. This is followed by a squeeze and excitation (SE) block to selectively emphasize informative features. A global average pooling layer is applied to the output of the SE block to generate a feature vector, which is then passed through a fully concatenated layer with four neurons per sound source category (ambient sounds, dolphin sounds, biological click sounds and ship noise). This is followed by a softmax activation function to generate a probability distribution across the sound source categories. The results show that the models should incorporate elements of the soundscape to achieve the desired results. The model is pre-trained on the ImageNet database of 1000 classes. Transfer training is performed by adapting the final layer of the CNN. Another approach for training with small data sets is the use of Support Vector Machines (SVMs) and unsupervised learning.

### C. Support vector machine and Generalized Additive Models

The class of SVM-based classifiers is used for binary classification between clicks and noise segments. For the detection of foraging clicks with low SNR [70] has offered a class-specific support vector machine (CS-SVM). Results are demonstrated for the detection and classification of dolphin clicks, beaked whale clicks and sperm whale echolocations. First, an energy detector is used to recognize regions of interest (ROIs) containing possible click sounds. The ROIs are then analyzed for feature extraction, i.e., to analyze the acoustic features of the detected clicks. Extracted features are fed into the CS-SVM classifier. A noise variable threshold adapts to different noise levels, ensuring effective detection of clicks. The Auto-Grouper algorithm is used for detection verification. This algorithm groups click sequences based on their periodicity, helping to identify and classify marine mammal vocalizations.

### D. Unsupervised methods

The method in [17] is a comprehensive method for identifying and classifying odontocetes clicks. This method characterizes clicks by their spectral patterns, such as low amplitude peaks and broad main peaks. The unsupervised Chinese Whispers clustering algorithm is used to identify dominant signal types based on spectral distances. The clusters are manually inspected and compared across sites to identify recurring signal types, focusing on spectral shape, inter-click interval (ICI) distribution and self-similarity. The method also includes parameter tuning for clustering to balance temporal resolution with data manageability. The main assumption in developing this method is the constancy of the click's spectral features.

### E. Remaining Challenges for Machine Learning Detection

One of the biggest challenges we see in the application of machine learning algorithms for acoustic detection of biofauna transient signals is the robustness for different underwater environments and for noise instances such as from snapping shrimps and vessel cavitation radiated noises. Proposed solutions are usually only suitable for certain contexts and often struggle with the variability and unpredictability of different marine soundscapes. These include the presence of similar-sounding species that are not part of the target objects, anthropogenic noise and different acoustic properties of the ocean. Another challenge is that only a limited amount of labelled data is available for training many of these algorithms. This is critical for supervised learning approaches and is exacerbated when dealing with rare or less studied species.

Future research could explore several promising avenues to fill these gaps. One approach is to improve feature extraction techniques to better capture the unique acoustic characteristics of different species, even in noisy environments. This could involve attention networks for the application of deep learning which have shown potential for dealing with limited data in other areas. In addition, the development of semi-supervised or unsupervised learning models could alleviate the problem of scarce labelled data by exploiting the abundant but unlabelled acoustic recordings. Research could also focus on developing more robust algorithms that can adapt to different ocean conditions and different noise profiles. Collaborative efforts to share and annotate data by researchers around the world could significantly enrich the datasets available for training and testing and improve the accuracy and reliability of the models.

## V. DATA DEPENDED METHODS

We call data-dependent methods a family of approaches that derive their processes from the data itself, similar to machine learning algorithms, but do not include a learning phase. One such approach is the template matching approach, which identifies patterns or features in the data and matches them to a template of the target signal. Other data-driven approaches use adaptive filters and the page test. The following is an overview of such approaches for click detection.

### A. Template Matching

In [71] details of the Transient Research Underwater Detector (TRUD), algorithms are presented. This scheme detects and classifies echolocation clicks through the spectrogram correlation. The correlation is based on templates of click patterns of different species. TRUDs based on a General Wideband Pulse Detector (GWPD), which uses narrowband energy accumulation and compares it across different time samples to detect potential whale-like clicks. The detected clicks are then organized into pulse trains and their statistical properties are evaluated. The method combines single click analysis with pulse train, thereby overcoming noise transients.

The method presented in [72] is a probabilistic approach that uses the concept of sampling entropy (SE). The method starts with the selection of an embedded dimension and constructs an embedding vector for each point in the time series that is comprised of consecutive samples. The correlation sum is a normalized count reflecting how many pairs of states (represented as vectors in the reconstructed phase space) are similar to each other within a certain level of tolerance (distance), excluding comparisons of a state with itself. The detection metric of SE is based on the natural logarithm of the conditional probability that a data set that has repeated for d samples within a tolerance r will also repeat for d+1 samples. Since clicks increase the standard deviation of the ambient noise such that the SE approaches to zero, the conditional probability is a good metric to separate signals from noise. Since no assumption on the noise distribution is required, the method is robust for different types of noise, including overlapping dolphin whistles and ship engine noise, and does not require prior training. However, it is difficult to distinguish clicks from noise transients, such as those which originate from snapping shrimp.

Another method that relies on a priori information about the signal is presented in [73]. The method is geared for tracking odontocetes (toothed whales) using a refined generalized cross-correlation (GCC) to adapt to noise environments and detect echolocation clicks while estimating the Time-Difference-of-Arrival (TDOA). Noise suppression is achieved by accumulating echolocation clicks over longer intervals. This, in turn, requires a clustering procedure to manage multiple TDOAs. The latter involves clustering of parameters such as location and speed using a factor graph-based multi-target tracking (MTT). The sum-product algorithm is used for tracking in the TDOA range, with a second MTT for 3D tracking by combining TDOAs from different hydrophone pairs. The method assumes that the clicks are stationary over a time interval longer than the GCC length. This, however, may limit its applicability in complex marine environments.

### B. Adaptive filters

In a matched filter, a given signal is correlated with a known waveform (the template) of the target signal in order to obtain the energy of the signal and the gain during processing for noise cancellation. In [74], the detection of sperm whale clicks is achieved by a stochastic matched filter (SMF), which correlates the incoming signal with a template signal, taking into account the statistical properties of the noise and echoes. In the SMF, the SNR is maximized by identifying the eigenvector associated with the largest eigenvalue in a given matrix equation. The detection function uses the linear filter applied to a typically small data window that matches the average length of a sperm whale click to determine whether the sound in that segment is likely to be a whale click or just random sea noise. The template is created from an average of 1,000 whale clicks. In [75] a method for localizing individual pulse-like underwater sounds using an array of hydrophones is offered. The localization involves a matched filter with an adaptive threshold. For each detected pulse, a dynamic window is applied using the call itself as a template. This dual MF approach proves to be more accurate than using a single MF. However, the signal is assumed to be stationary, which may not be true in all underwater environments.

The method in [76] detects sperm whale clicks using a recursive time-varying grid filter. At the heart of the method is the normalized recursive exact least-square time-variant lattice filter, which dynamically adapts to the signal's changing properties. This filter projects the signal onto a subspace defined by its past values. This approach accounts for changes in the second-order statistics of the signal captured by time-varying Schur coefficients, rather than relying on amplitude alone. The method does not require prior knowledge of the arrival time, amplitude or shape of the click. The algorithm also includes a forgetting factor that helps it adapt to the non-stationary nature of whale clicks by controlling the influence of past signal values, making it particularly effective in noisy environments. However, need for precise parameter calibration, such as the forgetting factor, could be challenging. An alternative to clicks correlation is detection by examining deviations from the expected values of the matched filter using the Page test.

### C. Page Test

The authors in [77] offer a modification of the Page test for low SNR environments. The approach performs wavelet analysis to remove noise transients prior to the page test. The Page test, also known as the cumulative sum test, is a sequential analysis method that detects changes in a data sequence. The test is based on comparing the variance and sample mean of a set of data with the expected mean and variance of a normal distribution. If the difference between the sample values and the expected values is significant, the test rejects the null hypothesis that the data is normally distributed and concludes that a signal is present. The Page test can also be adapted to detect transients corresponding to a click sound. The method in [78] uses the Page test for detecting sperm whale clicks. In this method, the signal is modeled as a state series and a distinction is made between 'noise' (absence of a click) and 'signal' (presence of a click) states. The transition between these states is determined by a signal strength statistic relative to two predefined thresholds. A constraint forces a 'noise' state if it remains longer than expected in the 'signal' state. Signal strength statistics are derived from estimates of instantaneous signal and noise power. These are calculated from the envelope of the waveform, which in turn is calculated using the Hilbert transform. After identifying potential clicks using the Page test, these clicks are categorized using an SVM with a quadratic kernel. The system verifies the decision by measuring the IPI between successive pulses. However, setting the detection threshold proves to be a challenge for robust detection as noise conditions vary in different marine environments. This is where threshold adjustment can help.

The click detection in [79] is based on the Page test. Once a potential click is identified, the algorithm analyzes the waveform, frequency spectrum and its resonance frequency to verify detection. A set of features is then extracted for each potential click, including the click duration, peak/centroid frequency, bandwidths, pulse zero crossing rate (ZCR) variance and exponential fit quality. These features are fed into an SVM for discrimination of near-axis sperm whale clicks from off-axis clicks and other transient noises.

In [80] a multi-stage method for detecting sperm whale clicks is presented. The method starts with a page test for transient detection. This is followed by envelope detection using the Hilbert transform. Thresholds are adapted by the noise level estimation. The method also demonstrates the ability to separate between direct and reflected click pairs based on amplitude variations and time intervals, which is further used in the localization part of the method. The method has proven robust to background noise and is insensitive to subtle variations in click amplitude and interval. However, it assumes a structure for the patterns of clicks, which imposes potential limitations.

*D. Adaptive threshold*

Adaptive threshold allows adjusting the detector to temporal characteristics of the data. This is particularly useful when dealing with data that changes spatially or temporally, such as directional whale clicks. In the context of click sound recognition, an adaptive threshold was used in [81] by coupling energy and frequency features. The clicks are characterized by repetitive arrivals with constant or slowly varying repetition periods within an assumed boundary for the ICI for sperm whale clicks. The detector analyzes the peaks of the histogram of arrival time differences at each hydrophone in search for dominant separations within the ICI range. Detection is declared if the corresponding arrival times show a regularity within a certain tolerance and their number exceeds a minimum time-varying threshold. A "detection event" is confirmed if either at least two detection flags are triggered in the current 1-minute recording or if one detection flag is triggered in the current recording and at least one more in one of the two previous 1-minute recordings. The real-time operation of the system is designed to process multiple hydrophones in parallel.

In [82] an energy detector is applied for identifying frequency bins that exceed a predetermined, time-averaged power threshold. The threshold is empirically set above the noise floor. Detection is declared by requiring parallel detection in a number of bins. This approach is particularly useful to detect broadband, impulsive signals such as clicks of sperm whales. In [83], a detection scheme based on the signal's kurtosis is offered to identify the expected sharpness in the samples' distribution in the case of a transient. The threshold is adapted to a sliding window to manage temporal heavy tail distributions when noise transients occur. However, the scheme is sensitive to scenarios with overlapping clicks. To handle such cases, another option for adjusting the threshold is spectral analysis.

For separating overlapping groups of echolocation clicks the frequency spectra, peak-to-peak amplitude, and IPI levels are used. In [84] the correlation of frequency spectra between clicks is used as a grouping metric while assuming that the characteristics of clicks from the same animal change gradually. To address the challenge of incorrectly classifying background noise as echolocation clicks, the algorithm uses the Low Percentage Removal Limit (LPRL) parameter, which is a critical component in the Click Group Separation algorithm, addressing the issue of background noise misclassification. Operating under the premise that falsely classified noise peaks constitute only a minor fraction of detected click groups due to their inconsistent Inter-Click Interval (ICI), amplitude, and frequency spectra, LPRL begins with a 0% setting. During the initial phase, an operator manually discerns and adjusts the LPRL to 1% above the percentage of total clicks identified as noise-related false positives. Subsequently, the algorithm is re-run, discounting "clicks" from groups below the set LPRL threshold. This procedure ensures the retention of authentic ICI values and amplitude thresholds, pivotal for precise click grouping. The introduction of LPRL significantly enhances the CGS algorithm's accuracy by effectively filtering out noise and reducing misclassification of echolocation clicks, particularly beneficial in noisy environments. Furthermore, the provision for manual adjustment of LPRL imparts flexibility to the algorithm, allowing it to be tailored to the unique noise characteristics of different datasets, thereby extending the CGS algorithm's applicability and robustness across diverse research settings. The algorithm adapts to the characteristics of the detected clicks.

Another form of adaptive processing is for the spectral energy. In [85] the adaptive Teager energy operator (A-TEO) is combined with an adaptive matched filter. The authors use adaptive windowing to account for the time-varying characteristics of the signals, and smoothing windowing to remove signal peaks that arise due to interference. The threshold is adjusted based on the mean and variance of the signal. In comparison with the TEO, the rainbow click detector and the spectral density (SD) detector shows an advantage attributed mostly to the smoothing processing. The proposed method relies on an accurate estimation of the ICI, which can be challenging at low SNR. A similar approach is used in [86], where the application of TKEO is used in combination with moving average filters. The TKEO output is further processed by two short moving average filters, a scaled Gaussian function (MAF1) and a rectangular averaging filter (MAF2), to provide near instantaneous spike detection. The filter difference ratio (FDR), a normalized measure derived from the outputs of MAF1 and MAF2, is critical for identifying spikes corresponding to clicks. This approach was developed to amplify the energy peaks corresponding to clicks while suppressing the harmonic components. The method has low computational complexity and can be used in real time.

The method in [87] uses spectrogram analysis. The algorithm uses a per-frequency bin, a dynamic threshold, which tests the multiplicative factor k over the exponential average of the power in the frequency bin. The result is a binary-valued spectrogram, where each bin exceeding the threshold is marked. As the algorithm processes the full bandwidth, it is able to capture a wide range of vocalization frequencies. In [88], a method for analyzing dolphin vocalizations in the presence of background noise such as boat propellers and engines is presented. The method for detecting peaks involves thresholding the SNR for suspected clicks. Detected peaks are classified based on their ICI to distinguish click sequences from noise transients. The classification filters out reverberation and overlapping clicks and focuses on identifying different click sequences by an adaptive ICI threshold.

### E. Remaining Challenges for Data depended methods

A main challenge we observe for data-driven techniques is in the effective processing of time-varying signals. Strong background noise, complex oceanic soundscapes and varying SNR can significantly affect the accuracy of species identification and detection performance. Another critical issue is to measure the IPI and ICI. Signal reverberations and overlapping sources can easily be mistaken as noise transients leading to misdetections. This will also occur when the animal's clicks are directional, causing time-variation in the SNR, which is often neglected when searching for constant sequences of pulses. A formal representation of the problem as a constraint optimization problem can assist in the rigorous analysis of the signals.

## VI. USEFUL INFORMATION ABOUT METHODS SURVEYED

### A. Available databases of clicks

An important part of our survey is a list of databases including echolocations that are openly available for testing. Several projects have kindly released their collected data. These datasets can serve as benchmarks to compare detection performance on a common basis, and to train in case of learning schemes. Publicly accessible and free databases we found are catalogued in Table I. The name of the dataset and a link to access it are given in the first two columns. We also list the location where the data was collected. The data type and the data size are listed, as well as an indication whether the data is labelled or not. There are additional available datasets that were not used in surveyed papers, collection of which can be found on [89] or [90].

### B. Open source click detection methods

As most of the methods we surveyed are complex to implement, several papers as well as software platforms share their implementation code. This is very useful for users or as benchmark schemes. PAMGUARD [107] is a semi-automated open-source software framework for passive acoustic detection and classification of cetaceans (whales, dolphins and porpoises). PAMGUARD serves as a platform for various passive acoustic detection techniques and tools that were previously available to provide a solution for researchers and users in the field. The platform is flexibly designed to process data from multiple sensors in any configuration. The software is highly modular, meaning it can be customized for different sensing platforms, e.g., offline processing of audio files or work in a real time mode. The supported vocalizations range from low-frequency moans to ultrasonic echolocation clicks. The user can choose between detection methods. In particular, the software uses a combination of pre-filters and trigger filters for click detection, optimizing the detection in the frequency band of interest and creating short sound clips for further analysis. Several parameters are required for system operation. These include the frequency, pattern, and intensity of the clicks produced by the target species. Performance of PAMGUARD is often used as an benchmark e.g., [86] [108], [109].

Another popular open source solution is Ishmael [110]. Ishmael offers several methods for marine bifauna passive detection, including energy detection, matched filtering and spectrogram correlation. The processing is performed over the signal's spectrogram. Detection thresholds are adaptively set such that fewer parameters are required from the user. Detection of signal sequences is also offered, which increases its usefulness for monitoring biological or mechanical sources with cyclostationary patterns. Both real time and offline modes of operations are possible. However, the method assumes a certain level of user knowledge in interpreting and customising the detection function, which could be a limitation for less experienced individuals. Research papers that compare performance with Ishmael are [111] and [112].

The Triton software package [113] serves as a platform for analyzing acoustic data. It offers the user a choice between click and whistle detection, and uses detection features such as power spectra, spectrograms and Long Term Spectral Averages (LTSA). The latter is efficient in condensing large data sets for display and analysis. Triton operates via MATLAB and offers a user-friendly graphical user interface that enables efficient review of large data sets. It offers functions such as reading raw data from the High-Frequency Acoustic Recording Package (HARP) and converting it into .xwav or .wav files. Users can interactively navigate through time series, spectrograms and spectra of single and multi-channel files. It also provides the ability to create and interact with LTSAs from a collection of files, facilitating long-term monitoring and detailed investigation of specific acoustic events. In addition, Triton supports data management by decimating high sample rate files for easier analysis of low-frequency sounds and saving data in different formats. An important aspect of Triton is its extensibility through remoras, which are user-developed MATLAB routines that can be integrated into Triton. This allows users to customize the software to their specific needs without changing the core functionality. Triton is used as benchmark in [114].

### C. Comparison of key algorithms

As part of our review, we implemented a number of methods and compared their performance. The methods were selected as representatives to the categories presented in the survey. Results are shown for the openly available PAMGUARD (the "click detector" module) and Triton (the "SPICE detector" remora) platforms, the method in [86] that applies an adaptive threshold, the detection scheme in [66] that employs a neural network, and the method in [85] for another representation of the adaptive threshold approach.

| DETECTION OF CLICKS: SUMMARY OF DATABASES | | | | | |
|---|---|---|---|---|---|
| archive name | link | collection site | tagged? | data type | number of recordings |
| MobySound | [91] | various studies on marine mammals | Yes | .wav | 14,000 vocalizations from eight species of baleen whales |
| Fisheries- Oceanography Coordinated Investigations | [92] | Bering Sea | Yes | .flac | different depending on the recording session |
| Cal-COFI Marine Mammal Data | [93] | California, USA | Partially | .wav | different depending on the recording session |
| Watkins Marine Mammal Sound Database | [94] | wide range of geographic areas | Yes | .wav | different depending on the recording session |
| DECAF - AUTEC Sperm Whales - Multiple Sensors - Complete Dataset | [95] | Atlantic Undersea Test and Evaluation Center (AUTEC) in the Tongue of the Ocean, Bahamas | Yes | .wav | 675 specimens |
| Orcasound - bioacoustic data for marine conservation | [96] | US/Canada | Yes | .wav | Live |
| DeepAL fieldwork data 2017/2018 | [97] | Northern British Columbia (Vancouver Island) | Yes | .wav | 31,928 audio clips; 5,740 (18.0%) killer whale and 26,188 (82.0%) noise labels. |
| Voice in the sea | [98] | All over the world | Yes | .wav | 31 cetacean and 12 pinnipeds |
| Dosits | [99] | All over the world | Yes | .wav | 30 Baleen Whales, 33 Toothed Whales, 25 Pinnipeds and 11 Sirenians |
| NOAA fisheries | [100] | Hawaii, USA | Yes | WAV | varies differently depending on the species |
| DCLDE 2022 Raw Passive Acoustic Data | [101] | Hawaii, USA | Yes | .wav .flac | least represented species 2, most more than 10000 |
| Zenodo | [102] | Ionian Sea | Yes | .png | 7,977 Files |
| SABIOD | [103] | south of Port-Cros National Parc / Cote Azur | No | .wav | 11 recordings of different lenght |
| Ocean glider observations in Greater Cook Strait, New Zealand | [104] | the Greater Cook Strait shelf sea, New Zealand | No | .nc | 7 recordings of different length |
| CIBRA of the University of Pavia | [105] | the Greater Cook Strait shelf sea, New Zealand | No | .wav and .mp3 | 14x2 recordings (in both formats) |
| The Dominica database | [106] | the Dominica island | Yes | .wav | 39 files of 5-minute recordings with sperm whale clicks and 43 files of 5-minute recordings with ambient sounds, ship noise and dolphin clicks and whistles. |

TABLE I: Table of publicly available databases and key characteristics.

The detection method in [86], referred to as *Gabor*, applies the TKEO over Gabor-transformed signals. Low complexity makes the approach suitable for online scenarios. The method was cited by 23 papers, and offers an effective usage of the TKEO for transient detection.

The detection method in [66], referred to as *statistical*, introduces a statistics-based approach for the identification of sperm whale clicks, primarily echolocation clicks and creaks. This method comprises statistical analysis of features, presence detection via a neural network, and classification of individual echolocation clicks and creaks. The paper has been cited so far by 7 papers, and its network architecture is well described, making it easy to implement.

The method presented in [85], termed *ATEO*, employs an adaptive energy-detector using the Adaptive Teager Energy Operator (ATEO), which utilizes a windowing technique to accommodate the time-varying characteristics of acoustic signals. The adaptive detection threshold offers robustness in different marine environments with little parameter calibration. The method was cited by 7 papers so far.

We implemented the above schemes and tested their performance on the AUTEC dataset [95]. This dataset includes 1364 manually annotated clicks, and was collected in the Tongue of the Ocean (Bahamas) using a single hydrophone deployed for 44 days. We chose this dataset since it provides low noise recordings, on top of which more noise can be synthetically added to test performance in varying SNR. Results are shown in Fig. 3 in terms of the detection vs. false alarm rates, and in Fig. 4 in terms of the detection rate as a function of the SNR. We observe that the best results are obtained with the statistical method in [66], but only at high SNR values. We report that the method is also easy to implement and only a few parameters need to be adjusted. The adaptive threshold method ATEO in [85] also requires only a few parameters for calibration. This method can work well at low SNR, but can only detect the presence of a single whale. The detection method implemented in the Triton platform provides good results at high SNR. Its advantage lies in its ease of use with a user-friendly GUI, but it requires many parameters for calibration and is more suitable for processing large files. Compared to the other methods, the Gabor method in [86] provides low results. However, it is easy to implement and relatively robust. Finally, PAMGUARD, with its "click detector" module, achieves good results even at low SNR, but is not robust based on the trade-off between detection and false alarm. Nevertheless, it provides a simple analysis platform that can operate in real-time, and its main advantage is the ability to process arrays of hydrophones to provide an angle of arrival estimate using a phased array.

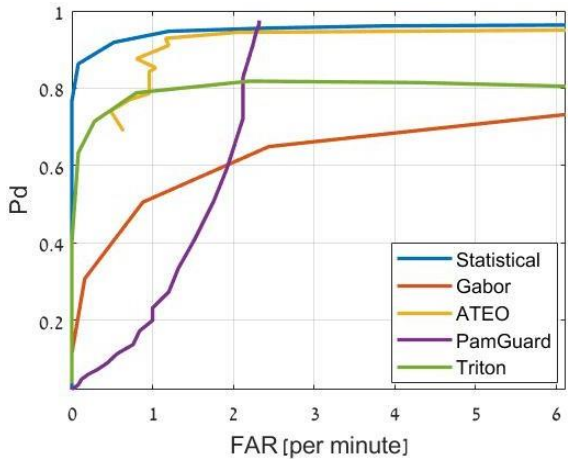

Fig. 3: Detection rate (Pd) vs. false alarm rate (FAR) for 5 click detection schemes over the AUTEC dataset [95].

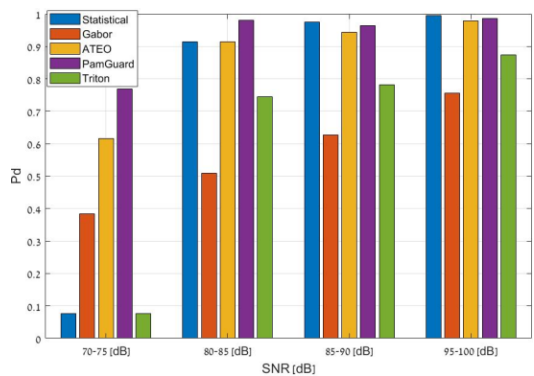

Fig. 4: Detection rate (Pd) vs. fsignal to noise ratio (SNR)for 5 click detection schemes over the AUTEC dataset [95].

### D. Summary of Click Detection algorithms

In this subsection we summarize the detection schemes surveyed in this paper. Table II presents four main challenges that were considered in the development of click detection methods, and lists the papers that directly handle these challenges. This table can support future research by directing authors to papers most relevant to their focus field. In Table III we group detection methods based on the type of data that was used for performance evaluation. Since most of the methods used a real dataset, the table further shows the public availability of the data. While half of the researched literature embraces openness, offering access to their data, the other half withholds their datasets from the public. This disparity not only hinders the validation and reproducibility of scientific findings but also stifles innovation. The absence of shared data curtails the potential for collective advancement, as researchers are deprived of the opportunity to build upon existing work, explore new hypotheses, or apply advanced analytical techniques to rich, pre-existing datasets. In this scenario the scientific community, and ultimately the research itself, loses the most. In Table V we identify common processing tools used by the detection methods. The tools are also divided by the approaches selected as subsections in our survey, and allow the reader to identify the type of analysis required when coming to detect clicks. In Table IV, methods are grouped by their application. This division can assist the choice of benchmark based on the considered scenario, e.g., real time analysis or offline processing of many files. The supervised and unsupervised labels used in this table refer to the need for manual labeling. In Table VI, we divide the detectors by the detection features that are used. Some of the features are used more frequently, while other features are used in only one research study. This list reflects the commonalities of clicks' attributes.

| CHALLENGES CONSIDERED IN THE LITERATURE | |
|---|---|
| **Challenges** | **Related Literature** |
| Low signal-to-noise ratio | [33], [37], [38], [43], [30], [52], [53], [61], [62], [63], [65], [66], [67], [68], [64], [69], [70], [74], [78], [83], [84], [85], [81], [87], [79], [82], [50], [31], [32], [34], [35], [41], [42], [45], [55], [56], [17], [72], [75], [76], [77], [80], [86], [88] |
| Time-varying noise | [33], [38], [61], [63], [65], [67], [64], [70], [74], [78], [83], [85], [81], [79], [82], [31], [41], [42], [51], [56], [17], [72], [73], [76], [86] |
| Simultaneous detection of multiple targets | [33], [37], [38], [39], [43], [49], [52], [61], [63], [68], [64], [69], [70], [74], [78], [84], [81], [87], [79], [82], [50], [34], [40], [51], [55], [56], [17], [71], [72], [73], [86], [88] |
| Non-stereotyped clicks | [39], [61], [70], [84], [85], [87], [82],[32], [33], [34], [35], [40], [41], [51], [56], [17], [73], [76], [86] |

TABLE II: Table of challenges the detection methods overcame.

| Data Source for PERFORMANCE EVALUATION | |
|---|---|
| **Evaluation** | **Related Literature** |
| Real data - data not shared | [37], [39], [44], [49], [63], [67], [74], [78], [84], [81], [87], [79], [35], [36], [45], [47], [55], [72], [75], [76], [77], [88] |
| Real data - data available publicly or on demand | [30], [33], [46], [52], [62], [65], [66], [68], [69], [70], [83], [85], [50], [31], [32], [33], [40], [42], [51], [53], [56], [71], [80] |
| Combining real and synthetic data | [38] , [43], [61], [64], [34], [41], [17], [73], [86] |

TABLE III: Table of types of data the detection methods were tested on.

| APPLICATIONS | |
|---|---|
| **Application** | **Related Literature** |
| (near) Real-time | [33], [38], [39], [43], [52], [61], [66], [69], [70], [74], [85], [81], [87], [79], [82], [31], [32], [33], [40], [41], [51], [53], [17], [71], [72], [75], [76], [86] |
| Offline | [33], [37], [44], [30], [46], [49], [30], [62], [63], [65], [67], [68], [64], [78], [83], [84], [50], [34], [35], [36], [42], [45], [47], [55], [56], [73], [77], [80], [88] |
| Supervised | [39], [44], [30], [46], [30], [52], [61], [62], [63], [65], [66], [67], [68], [64], [69], [70], [87], [40], [42], [45], [47], [55], [56], [53], [71], [73], [77], [88] |
| unsupervised | [33], [43], [31], [32], [33], [34], [35], [36], [41], [51], [17], [72], [75], [76], [80], [86] |
| Available implementation | [33], [63], [65], [78], [88] |

TABLE IV: Table of possible method applications.

## VII. CONCLUSION

The importance of monitoring echolocation clicks is demonstrated by the need to analyze behavior changes, explore population changes, and evaluate environmental impacts of anthropogenic activities. In this survey we aimed to systematically categorize and evaluate a broad spectrum of methodologies for detecting cetacean echolocation clicks. We provided an overview of feature analysis, machine learning-based detection, and data-dependent methods. Feature analysis techniques delved into the intricate characteristics of clicks, such as duration, phase, frequency and energy, employing signal processing tools to separate clicks from the ambient noise. Machine learning-based detection emerged as a promising frontier, with methods like convolutional neural networks (CNNs) and recurrent neural networks (RNNs) offering pattern recognition capabilities. Data-dependent methods provided a structured approach to comparing signals against predefined templates, harnessing specific characteristics of the target clicks for detection. We have also surveyed datasets openly shared for performance evaluation, and open software platforms. To comment on the suitability of the different approaches, we implemented representative schemes and tested their detection performance over a single dataset. Despite the advancements and the diversity of approaches reviewed, it is imperative to recognize that no single technique currently suffices to detect and classify the vocalizations of all known cetacean species in a robust manner. This reality underscores some of the remaining challenges in the field. These challenges include dealing with the variability and unpredictability of marine soundscapes, the scarcity of labeled data for algorithm training, and the need for algorithms that are robust against environmental noise, shipping cavitation noise and interference from other marine fauna. Addressing these limitations calls for a multifaceted approach: enhancing feature extraction techniques, embracing the potential of deep learning while ensuring adaptability to limited data. A particular challenge lies in the separation of clicks from simultaneously emitting animals. Finally, we divided the works surveyed by their application, tools used, and application to serve for future development of click detection techniques.

| DETECTION OF CLICKS: SUMMARY OF THE MAIN TOOLS USED | |
|---|---|
| **Evaluation** | **Related Literature** |
| Clustering Methods | [46], [49], [78],[33] |
| Fourier Analysis | [38], [44], [30], [82], [51], [56], [71] |
| TK operator | [44], [52], [61], [63], [85], [33], [32], [33], [45], [53], [55], [56], [86] |
| Phase Slope Function (PSF) | [32], [33] [37] |
| Walvet Transformation | [38], [77] |
| Empirical Mode Decomposition (EMD) | [39], [43], [40] |
| Hilbert Transform (HHT) | [43], [40], [42] |
| The radial basis function (RBF) activation function | [30] |
| Convolution neural network (CNN) | [61], [62], [63], [65], [64], [69] |
| Gabor curve-fitting method | [61], [86] |
| Multilayer Perceptron (MLP) | [66], [67] |
| SVM | [70], [79] |
| (Stochastic) Matched Filter | [74], [75], [76] |
| Page test | [78], [79], [77] |
| Kurtosis | [83] |
| Complex Autocorrelation Function | [34] |
| CCWEEMDAN | [41] |
| Cross correlation | [73] |

TABLE V: Table of main tools utilized in click detection methods.

| DETECTION OF CLICKS: SUMMARY OF THE FEATURES USED FOR DETECTION | |
|---|---|
| **Features** | **Related Literature** |
| IPI | [33], [30], [63], [78], [79], [50] |
| ICI | [33], [38], [46], [49], [68], [84], [81], [79], [82], [31], [32], [33], [34], [35], [36], [42], [88] |
| TDoAs | [63], [81], [32], [33], [34], [73], [75] |
| Peak Amplitude | [47] , [84], [33], [82], [88] |
| Duration | [33], [44], [31], [76], [80] |
| Spectral Bandwidth | [49], [78], [84], [17] |
| Phase | [37], [56] |
| Energy | [38], [46], [52], [87], [82], [40], [41], [45], [51], [53], [55], [56], [17], [86] |
| Frequency | [43], [51], [56], [17], [71], [80], [86] |
| Standard deviation and dynamic range of energy | [66] |
| Average Cepstral Features | [67], [45] |
| Entropy | [72] |
| Not specified | [61], [62], [65], [64], [69], [74], [77] |

TABLE VI: Table of features used for click detection.